# Deep Domain-Adversarial Adaptation for Automatic Modulation Classification under Channel Variability

Khondakar Ashik Shahriar
*Department of Electrical and Electronic Engineering*
*Bangladesh University of Engineering and Technology*
Dhaka, Bangladesh
kh.ashikshahriar@gmail.com

*Abstract*— Automatic Modulation Classification (AMC) plays a significant role in modern cognitive and intelligent radio systems, where accurate identification of modulation is crucial for adaptive communication. The presence of heterogeneous wireless channel conditions, such as Rayleigh and Rician fading, poses significant challenges to the generalization ability of conventional AMC models. In this work, a domain-adversarial neural network (DANN) based deep learning framework is proposed that explicitly mitigates channel-induced distribution shifts between source and target domains. The approach is evaluated using a comprehensive simulated dataset containing five modulation schemes (BPSK, QPSK, 16QAM, 64QAM, 256QAM) across Rayleigh and Rician fading channels at five frequency bands. Comparative experiments demonstrate that the DANN-based model achieves up to 14.93% absolute accuracy improvement in certain modulation cases compared to a baseline supervised model trained solely on the source domain. The findings establish the engineering feasibility of domain-adversarial learning in AMC tasks under real-world channel variability and offer a robust direction for future research in adaptive spectrum intelligence.

## I. INTRODUCTION

In wireless communication systems, Automatic Modulation Classification (AMC) is becoming more and more crucial for tasks like dynamic spectrum access, interference detection, and signal recognition. These features are especially important for autonomous communication networks and cognitive radio systems [1]. Conventional AMC methods are typically developed under the presumption of stable or idealized channel conditions and frequently rely on manually created features. But in reality, wireless signals usually face different propagation conditions, like Rayleigh and Rician fading, which can cause distribution shifts that restrict how broadly a model can be applied [2],[3].

End-to-end learning of signal features has been made possible by recent developments in deep learning, which has enhanced AMC performance [4],[5]. Deep models may still be susceptible to channel variability in spite of these advancements, particularly if they are trained and assessed using data from various fading environments. This problem, also known as a domain shift, makes it difficult to deploy AMC systems in real-world environments where precise channel characteristics might not be known beforehand.

In order to address this issue, the current study looks into how to mitigate performance degradation brought on by channel-induced domain shifts by using a domain-adversarial neural network (DANN) [6]. This method seeks to encourage feature representations that are less susceptible to the underlying channel conditions by introducing a domain classifier with adversarial training. By doing this, the technique aims to enhance AMC generalization without the need for labeled samples from the target channel domain.

A publicly accessible dataset [7] with five modulation schemes (BPSK, QPSK, 16QAM, 64QAM, and 256QAM) simulated under Rayleigh and Rician fading conditions across various frequency bands is used in this study. With an emphasis on two domain adaptation directions, Rayleigh to Rician and Rician to Rayleigh, the analysis compares the suggested DANN approach with traditional deep learning baselines. In addition to performance metrics and statistical testing, visualization tools like confusion matrices and t-SNE plots are used to evaluate the usefulness of domain-adversarial learning in this situation. This work's main contributions are as follows:

- An examination of domain-adversarial learning for AMC in the presence of channel fluctuations;
- empirical assessment in various channel adaptation settings and modulation types;
- analysis of interpretability and visualization to enhance comprehension of feature space alignment;
- To determine the significance of observed improvements, statistical validation is used.

Although the findings imply that DANN can provide resilience to specific domain shifts, the study also points out drawbacks and new research directions, especially concerning modulation-specific sensitivity and model complexity.

## II. LITERATURE REVIEW

AMC is a key component in modern communication systems, enabling adaptive demodulation, interference detection, and spectrum monitoring without prior knowledge of the modulation format. Early AMC approaches were largely likelihood-based or feature-based. Likelihood-based classifiers, such as those proposed by Azzouz and Nandi [8], modeled the received signal statistics under various hypotheses and used maximum likelihood detection. Although optimal under known channel conditions, their computational complexity and sensitivity to channel estimation errors limited practical deployment. Feature-based methods, in contrast, extracted handcrafted features from time, frequency, or higher-order statistics (e.g., cumulants [9], cyclostationary features [10]) before applying classical classifiers such as k-nearest neighbors (KNN), support vector machines (SVM), or decision trees. While computationally lighter, these methods suffered degraded performance under low signal-to-noise ratio (SNR) or severe channel fading. With the rise of deep learning, data-driven AMC methods began to outperform handcrafted pipelines by directly learning discriminative features from raw I/Q samples or spectrogram representations. Convolutional Neural Networks (CNNs) [11,12] effectively captured local temporal-frequency patterns, while Recurrent Neural Networks (RNNs) and Long Short-Term Memory (LSTM) networks [13] modeled temporal dependencies in modulation sequences. Hybrid CNN–LSTM architectures further improved robustness under varying SNR conditions [14]. More recent works incorporated attention mechanisms [15] and residual networks [16] to enhance feature representation. However, the majority of these studies assumed that training and testing data were drawn from the same distribution, which rarely holds in real-world wireless environments.

Domain adaptation methods address the domain shift problem by aligning feature distributions between a labeled source domain and an unlabeled target domain. The DANN framework proposed by Ganin et al. [17] introduced a gradient reversal layer to encourage domain-invariant feature learning. The present work builds upon these foundations by evaluating deep DANN-based AMC under Rayleigh-to-Rician and Rician-to-Rayleigh fading conditions at multiple sampling frequencies, providing both quantitative and qualitative analysis of domain alignment effectiveness.

## III. METHODOLOGY

### A. Theoretical Motivation

A core challenge in machine learning problems is the domain shift problem, where the data distribution encountered during deployment differs from that used in training and fine-tuning [18]. The theory of domain adaptation establishes that the error on a target domain is bounded by three factors:

1. The classification error in the source domain,
2. The divergence between the source and target feature distribution, and
3. The minimal joint error achievable by any ideal classifier across both domains.

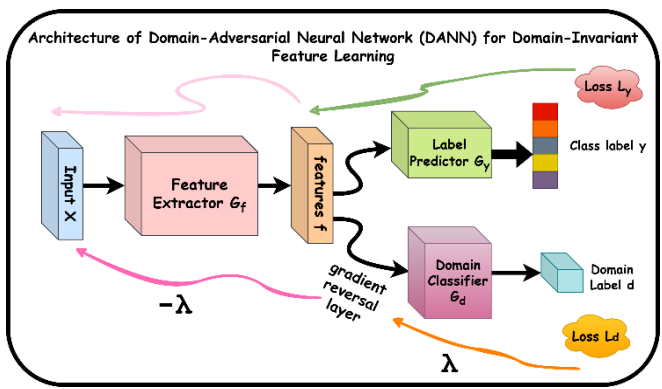

*Figure 1: Architecture of DANN, where a feature extractor produces a latent representation optimized for modulation classification and domain invariance with GRL and domain loss.*

Minimizing the divergence term is therefore critical to achieve robust generalization across domains. DANN framework provides a principled solution by promoting domain-invariant representation learning. This is done by two competing objectives:

- A label predictor is trained to correctly classify samples based on features extracted from the input.
- A domain classifier is trained to identify whether a sample originates from the source or target domain.

A Gradient Reversal Layer (GRL) connects the feature extractor to the domain classifier, reversing the gradient during backpropagation. This forces the feature extractor to learn features that reduce the domain classifier's ability to discriminate between domains, thereby aligning the feature distributions in the latent space.

Figure 1 illustrates the process, where the input x is first processed via feature extractor $G_f$, generating intermediate features f. These features are passed in parallel to the label predictor $G_y$ and domain classifier $G_d$. The label predictor minimizes the classification loss $L_y$, encouraging features to retain task-relevant information only. Meanwhile, GRL scales the domain classification gradient by a negative factor -λ, feature extractor to learn domain-invariant features by maximizing the domain classifier's loss $L_d$. The joint optimization is done by:

$$E(\theta_f, \theta_y, \theta_d) = \frac{1}{n}\sum_{i=1}^{n} L_y^i(\theta_f, \theta_y) - \lambda \left(\frac{1}{n}\sum_{i=1}^{n} L_d^i(\theta_f, \theta_d) + \frac{1}{n'}\sum_{i=n+1}^{N} L_d^i(\theta_f, \theta_d)\right) \quad (1)$$

Where $\theta_f, \theta_y, \theta_d$ represents the parameters of the feature extractor, label predictor, and domain classifier, respectively, and λ controls the trade-off between classification accuracy and domain invariance.

In the context of AMC under channel variability (e.g., Rayleigh vs. Rician fading), this loss formulation is particularly effective. The classification loss Ly ensures that modulation-specific characteristics are preserved, enabling accurate signal identification. The domain loss $L_d$, combined with the GRL, suppresses channel-specific artifacts that would otherwise bias the model towards the source channel. As a result, the model can generalize better to unseen channel conditions, maintaining stable performance in practical wireless environments.

### B. Dataset Description

Five digital modulation schemes, BPSK, QPSK, 16-QAM, 64-QAM, and 256-QAM, that were simulated under Rayleigh and Rician fading channels at sampling frequencies of 1 MHz, 10 MHz, 100 MHz, 500 MHz, and 1 GHz, are represented in the dataset using feature extraction. All channel-frequency combinations are covered by ten CSV files, and MATLAB scripts are included for reproducibility. 202 features are used to describe each sample. These features are obtained from (i) time and frequency domain signal processing metrics (statistical, spectral, and signal-specific measures); (ii) spectrogram-based features from STFT representations (statistical, spectral, and amplitude-based attributes); and (iii) image-based features from spectrograms using BRISK, MSER, and GLCM methods. This high-dimensional dataset serves broader purposes in spectrum sensing, cognitive radio, feature selection, and dimensionality reduction benchmarking. It is also well-suited for AMC under channel variability.

## C. Data Preprocessing

Five digital modulation schemes (BPSK, QPSK, 16-QAM, 64-QAM, and 256-QAM) were studied using feature-extracted numerical datasets at five sampling frequencies (1 MHz, 10 MHz, 100 MHz, 500 MHz, and 1 GHz) across Rayleigh and Rician fading channels. The first step in cleaning each dataset was to ensure consistent floating-point representation and convert complex-valued entries to their absolute magnitudes. StandardScaler was used to standardize features in order to enforce unit variance and zero mean. In order to prevent information leakage, the scaler was applied to both the source and target domains and only fitted on the source domain. Each of the five modulation classes had an integer-encoded label. The source data was split into 80% training and 20% validation for every pair of source and target domains, and the target data was equally split into unlabeled training.

## D. Model Architecture

Two different architectures were used:
- The baseline classifier is a shallow multiplier perceptron (MLP) that has an input layer matching the feature dimension, two fully connected hidden layers (256 and 128 units) with regularization (Batch Normalization, ReLU activation, and dropout), and a final output layer of size 5 for classification.
- Three fully connected layers (512, 256, and 128 units) make up the DANN, a deeper feature extractor that uses dropout, GeLU activation, and Batch Normalization. The extracted features are then passed to a label classifier (two fully connected layers with ReLU activation) to predict the modulation class, a domain classifier (two fully connected layers with ReLU activation) preceded by a GRL to encourage domain-invariant feature learning.

## E. Training Procedure

The models were implemented in PyTorch and optimized using the Adam optimizer with an initial learning rate of $1\times10^{-4}$ and a batch size of 128. The domain loss weighting factor $\lambda$ was gradually increased from 0 to 1 using a sigmoid schedule to stabilize training in the early epochs. Training was performed for 50 epochs with early stopping based on validation accuracy in the target domain. For each sampling frequency, experiments were conducted in both transfer directions: Rayleigh to Rician and Rician to Rayleigh.

## F. Evaluation Metrics

The model's performance was evaluated using multiple complementary metrics. Classification Accuracy (CA) measured the proportion of correctly predicted target domain samples, while Average Classification Accuracy (Avg. Acc.) quantified the mean accuracy across all modulation classes. To assess the benefit of domain adaptation, Absolute Improvement and Percentage Improvement were computed as the difference and relative gain, respectively, between the DANN and baseline models. Domain Classification Accuracy (DCA), measured before and after adaptation, provided insight into the degree of feature space alignment between domains, with lower post-adaptation DCA indicating improved domain invariance. Additionally, t-SNE visualizations were generated to qualitatively examine changes in feature separability and domain overlap resulting from adaptation.

## IV. EXPERIMENTAL RESULTS

The experimental results demonstrate that the effect of DANN in mitigating between Rayleigh and Rician fading channels across five sampling frequencies.

### A. Quantitative analysis

Table 1 shows the DCA before and after adaptation. At lower and mid frequencies (1–100 MHz), DANN achieves small but consistent improvements, indicating reduced domain discrepancy. For example, at 100 MHz, DCA increased from 0.7549 to 0.7574 for Rayleigh to Rician and from 0.7380 to 0.7454 for Rician to Rayleigh. At high frequencies (500 MHz and 1 GHz), DCA remained at 1.0, suggesting perfect domain separability and minimal benefit from adaptation.

**Table 1.** Domain Classification Accuracy (DCA) before and after adaptation.

| Frequency | Direction | DCA Before | DCA After |
|---|---|---|---|
| 1 MHz | Rayleigh→Rician | 0.7400 | 0.7303 |
| 1 MHz | Rician→Rayleigh | 0.7320 | 0.7269 |
| 10 MHz | Rayleigh→Rician | 0.7237 | 0.7351 |
| 10 MHz | Rician→Rayleigh | 0.7220 | 0.7254 |
| 100 MHz | Rayleigh→Rician | 0.7549 | 0.7574 |
| 100 MHz | Rician→Rayleigh | 0.7380 | 0.7454 |
| 500 MHz | Rayleigh→Rician | 1.0000 | 1.0000 |
| 500 MHz | Rician→Rayleigh | 1.0000 | 1.0000 |
| 1 GHz | Rayleigh→Rician | 1.0000 | 1.0000 |
| 1 GHz | Rician→Rayleigh | 1.0000 | 1.0000 |

Table 2 presents average baseline and post-adaptation accuracies. At 100 MHz, the framework improved average accuracy by +2.94% (Rayleigh to Rician) and +2.46% (Rician to Rayleigh). Similar trends are observed at 1 MHz and 10 MHz, with improvements of up to +4.10%.

**Table 2.** Average classification accuracy (%) before and after adaptation, with absolute and percentage improvements.

| Frequency | Direction | Avg. Baseline | Avg. DANN | Abs. Improvement | % Improvement |
|---|---|---|---|---|---|
| 1 MHz | Rayleigh→Rician | 76.86 | 79.33 | +2.47 | +3.21 |
| 1 MHz | Rician→Rayleigh | 74.94 | 74.69 | −0.25 | −0.33 |
| 10 MHz | Rayleigh→Rician | 82.78 | 83.54 | +0.76 | +0.92 |
| 10 MHz | Rician→Rayleigh | 74.65 | 77.71 | +3.06 | +4.10 |
| 100 MHz | Rayleigh→Rician | 85.57 | 88.08 | +2.52 | +2.94 |
| 100 MHz | Rician→Rayleigh | 82.41 | 84.44 | +2.02 | +2.46 |
| 500 MHz | Rayleigh→Rician | 29.90 | 20.00 | −9.90 | −33.12 |
| 500 MHz | Rician→Rayleigh | 37.64 | 34.03 | −3.61 | −9.60 |
| 1 GHz | Rayleigh→Rician | 41.26 | 20.00 | −21.26 | −51.53 |
| 1 GHz | Rician→Rayleigh | 20.00 | 20.00 | 0.00 | 0.00 |

However, performance deteriorates at 500 MHz and 1 GHz, with drops exceeding 50% in some cases, indicating that adaptation struggles under severe channel mismatch at higher frequencies.Table 3 provides the modulation-wise performance for 100 MHz. For Rayleigh to Rician, the largest gain was in BPSK (+3.85%), while QPSK and 16-QAM saw reductions. For Rician to Rayleigh, BPSK achieved a significant +14.93% improvement, and 256-QAM gained +2.37%, but QPSK and 64-QAM declined slightly. These findings suggest that DANN particularly benefits lower-order modulation schemes under moderate domain shift.

**Table 3.** Per-class accuracy (%) before and after adaptation for 100 MHz.

| Direction | Modulation | Baseline | DANN | Δ Accuracy |
|---|---|---|---|---|
| **Rayleigh→Rician** | BPSK | 84.38 | 88.24 | +3.85 |
| | QPSK | 92.49 | 88.84 | −3.65 |
| | 16-QAM | 58.05 | 47.71 | −10.34 |
| | 64-QAM | 88.91 | 89.49 | +0.58 |
| | 256-QAM | 100.00 | 100.00 | 0.00 |
| **Rician→Rayleigh** | BPSK | 55.01 | 69.94 | +14.93 |
| | QPSK | 99.41 | 96.85 | −2.56 |
| | 16-QAM | 95.17 | 95.17 | 0.00 |
| | 64-QAM | 64.62 | 60.19 | −4.42 |
| | 256-QAM | 87.77 | 90.14 | +2.37 |

*B. Qualitative Analysis*

To further examine the impact of DANN on feature representation, t-SNE visualizations of the learned embeddings were generated for selected frequencies. At 100 MHz (Figure 2), DANN adaptation yields more compact and well-separated clusters across modulation classes compared to the baseline, indicating improved domain alignment and enhanced discriminative capability. At 500 MHz (Figure 3), although DANN reshapes the feature space, cluster elongation and partial overlaps remain, consistent with the drop in quantitative performance. This suggests that while adversarial adaptation facilitates inter-domain feature alignment, extreme channel and sampling conditions may still impose limitations on the separability of modulation classes, highlighting the need for more robust feature extraction strategies or hybrid adaptation approaches.

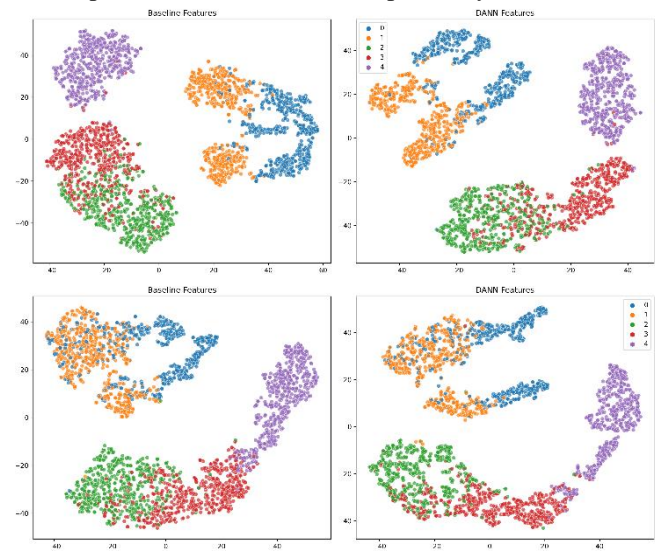

*Figure 1: t-SNE plots showing baseline vs. DANN adaptation for Rayleigh to Rician(top) and Rician to Rayleigh(bottom), with improved class separation under DANN*

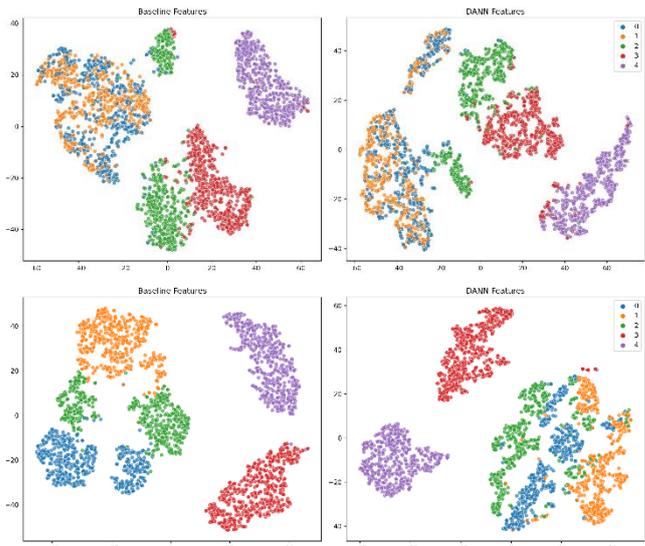

*Figure 3: t-SNE visualization of baseline and DANN features for 500 MHz, showing limited improvement in inter-class separation after domain adaptation*

## V. DISCUSSION

This study demonstrates that domain-adversarial training can moderately improve Automatic Modulation Classification performance under channel variability, particularly for source–target transfers between Rayleigh and Rician channels at 1–100 MHz. Across these conditions, the DANN approach achieved small but consistent gains in average accuracy for most domain pairs, with qualitative t-SNE visualizations indicating better clustering of modulation classes compared to the baseline. These improvements highlight the potential of feature-level domain alignment in mitigating moderate channel-induced distribution shifts.

However, performance gains were not uniform across all modulation types, with certain higher-order QAM schemes showing negligible or negative changes. Moreover, at 500 MHz and 1 GHz, both baseline and DANN accuracies dropped significantly, suggesting that the model's capacity to align domains weakens under severe channel distortion and noise. These findings indicate that while DANN can enhance robustness in moderate shift scenarios, it may require complementary techniques, such as signal preprocessing, channel compensation, or hybrid adaptation strategies—for more challenging conditions.

## VI. CONCLUSION

This work presented a DANN framework for improving AMC under channel variability between Rayleigh and Rician fading environments. By enforcing domain-invariant feature learning through adversarial training, the proposed method achieved moderate but consistent gains in classification accuracy for most source–target domain pairs at sampling frequencies up to 100 MHz. Qualitative analysis using t-SNE visualizations supported these improvements, revealing enhanced clustering and reduced overlap between modulation classes compared to a baseline model trained solely on the source domain.

Despite these benefits, performance gains were not uniform across all modulation formats, and substantial degradation persisted at higher frequencies (e.g., 500 MHz and 1 GHz), indicating reduced adaptation efficacy under severe channel distortions. These findings suggest that while DANN offers a promising direction for robust AMC in practical cognitive radio systems, future work should investigate hybrid adaptation strategies combining adversarial alignment with advanced signal preprocessing and channel compensation techniques to enhance generalization across more challenging channel conditions.

ACKNOWLEDGMENT

The author would like to thank the maintainers of the open-source modulation classification dataset used in this study for making their work publicly accessible, enabling reproducible research. The computational resources and GPU support provided by Kaggle were instrumental in conducting the experiments.